\documentclass[floatfix,nofootinbib,superscriptaddress,twocolumn,pre,showpacs,showkeys,preprintnumbers]{revtex4-2}
\usepackage{amsmath,amssymb}
\usepackage{epsfig,graphics,color,calc,graphicx}
\usepackage{enumitem}
\usepackage{subfigure}

\usepackage{mathrsfs}
\usepackage[title]{appendix}
\newcounter{enmccommento}

\begin{document}
%%% Font: commenta la riga che segue se vuoi i font standard
%\fontfamily{ppl}\selectfont
%\logo
\title{Deterministic model of battery, uphill currents and non--equilibrium phase transitions}

\author{Emilio N.\ M.\ Cirillo}
\email{emilio.cirillo@uniroma1.it}
\affiliation{Dipartimento di Scienze di Base e Applicate per l'Ingegneria,
             Sapienza Universit\`a di Roma,
             Via A.\ Scarpa 16, I--00161, Roma, Italy.}
%\thanks{ENMC acknowledges Eurandom for the kind hospitality.}

\author{Matteo Colangeli}
\email{matteo.colangeli1@univaq.it; ORCID 0000-0002-7424-7888}
\affiliation{Dipartimento di Ingegneria e Scienze dell'Informazione e
Matematica,
Universit\`a degli Studi dell'Aquila, Via Vetoio, 67100 L'Aquila, Italy.}

\author{Omar Richardson}
\email{omar.richardson@kau.se}
\affiliation{Department of Mathematics and Computer Science,
Karlstad University, Sweden.}

\author{Lamberto Rondoni}
\email{lamberto.rondoni@polito.it; ORCID 0000-0002-4223-6279}
\affiliation{Dipartimento di Scienze Matematiche,
Politecnico di Torino, Corso Duca degli Abruzzi 24, I--10129 Torino, Italy}
\affiliation{INFN, Sezione di Torino, Via P. Giuria 1, 10125 Torino, Italy}

\begin{abstract}
We consider point particles in a table made of two circular cavities connected by two rectangular
channels, forming a closed loop under periodic boundary conditions. In the first channel, a bounce--back mechanism acts when
the number of particles flowing in one direction exceeds a given threshold $T$. In that case,
the particles invert their horizontal velocity, as if colliding with vertical walls. The second channel is divided in two
halves parallel to the first, but located in the opposite sides of the cavities. In the second channel, motion is free. We show that, suitably tuning the sizes of cavities, of the channels and of $T$, non--equilibrium
phase transitions take place in the $N\rightarrow \infty$ limit. This induces a stationary current in the circuit, thus modeling a kind of battery, although our model is deterministic, conservative, and time reversal invariant.
\end{abstract}



%\ams{00A71,  80M31, 60J60, 92D25, 35L65}

%\preprint{Appunti: \today}

\maketitle

%\vfill\eject
\section{Introduction}
\label{sec:intro}
\par\noindent

The nature of non--equilibrium phenomena is diverse and rich, and a theory encompassing them is still in 
the making \cite{Kubo91,EvMo,Gall14,Bertini15}. Such a task requires, in particular, understanding the 
coupling with external agents or reservoirs that may locally allow the condition of detailed balance 
or its violation \cite{Lebowitz,Maes,BePuRoVu,Conti2013,CdMP1}.
In this work we provide numerical results and a theory explaining the onset of stationary currents in deterministic conservative 
reversible systems made of $N$ point particles. Such currents are generated by non--equilibrium phase transitions, that result in
a deterministic model of battery, which is phase space volumes preserving and time reversal invariant \cite{BDL10}.
Flows and oscillations produced by this mechanism resemble those observed in biological systems or chemical reactions, cf.\
Refs. \cite{Kur84,Wil12} for classical and quantum oscillators, and Ref. \cite{Zha17} for experiments on time crystals. 
In particular, our single component deterministic model shows a realization of \textit{uphill currents}, {\em i.e.}\ currents opposing the driving fields, thus providing an instance of the so-called
{\it negative absolute mobility} \cite{Eich03,Ros05,Muk18}. A theoretical description of uphill diffusions was given in \cite{CdMP3,CGGV18}
for stochastic spin models coupled to external reservoirs; moreover, in \cite[Sec.4.5]{ACCG19} uphill currents were also obtained from the scaling limit of inhomogeneous random walks on a lattice.
A non--equilibrium phase transition occurring in a deterministic particle system was recently observed in a model with two cavities connected by a single channel, allowing no stationary currents \cite{CCMRR2020}. 
The transition amounts to switching from a homogeneous state, in which approximatively the same number of particles lies in each urn, to an inhomogeneous state in which almost all particles gather in a single urn. The model studied in \cite{CCMRR2020} was also amenable to a stochastic interpretation, in terms of time-dependent Markov chains. In this work, we investigate the nature of the steady state in a two-urns model equipped also with a second channel, that permits to close the system as in a circuit.
The main question we address here is two--fold. First, we shed light on the existence of non--equilibrium phase transitions for the circuit model, in which the second channel is designed to contrast the formation of particle gradients between the urns. Furthermore, we also discuss the emergence of stationary currents, flowing through the circuit and sustained by the phase transitions. We shall thus unveil a non--trivial phase diagram for our model, revealing that phase transitions indeed occur in certain regions of the parameter space and are always followed by stationary currents.

The work is organized as follows. In Sec. \ref{sec:sec1} we introduce our model and also present the numerical results of our deterministic dynamics.
In Sec. \ref{sec:sec2} we tackle the theoretical investigation of the model by means of probabilistic arguments, and also compare the theoretical prediction with the numerical results. We also highlight strength and limitations of the probabilistic model.
More details on the probabilistic derivation are deferred to the Appendix \ref{sec:app}.
Finally, conclusions are drawn in Sec. \ref{sec:sec3}.

\section{The model}
\label{sec:sec1}
\par\noindent

\begin{figure}[h]
%\centering
\includegraphics[width = 0.45\textwidth]{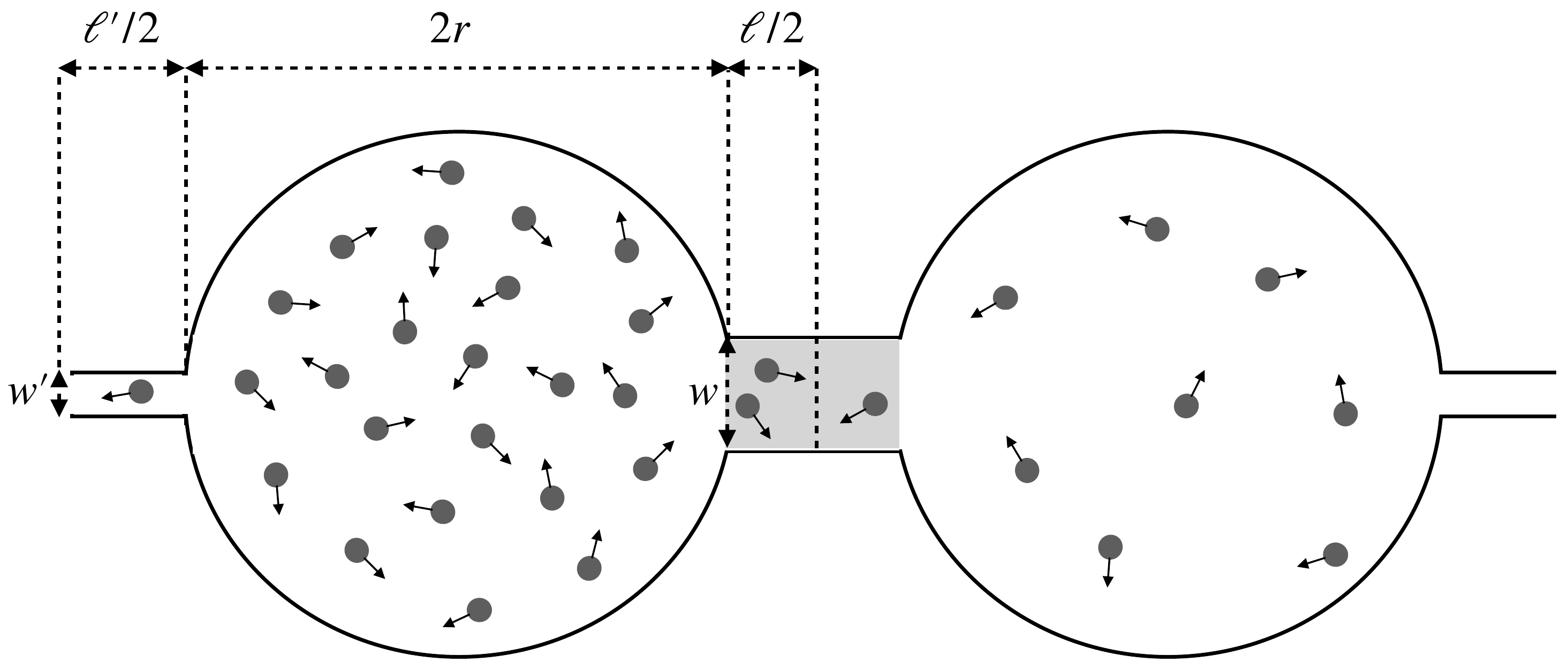}
\caption{The billiard table: the $N$ point particles
are represented as small disks and velocities are represented by
arrows. The two grey shaded regions, in the first channel, are the gates in which the bounce--back mechanism separately acts. The horizontal component of the velocity of the particles contained in one gate and moving toward the other is reversed whenever their number is larger than the prescribed threshold value $T$.}
\label{fig:model}
\end{figure}

Our model consists of $N$ point particles that move in straight lines with speed $v=1$,
and collide elastically with hard walls. Hence, from collision to collision, the particles follow these equations of motion: ${\bf \dot q}={\bf p}$ and ${\bf \dot p}={\bf 0}$. Therefore, their speed is preserved while their velocity
is reflected with respect to the normal to the boundary of the table, at the
collision point. The billiard table is made of two circular \emph{urns}
of radius $r$, connected by two rectangular channels of widths $w,w'$ and lengths
$\ell,\ell'$, called, respectively, first and second channel, cf.\ Fig.~\ref{fig:model}. The two urns will be referred to, in the sequel, as urn 1 and urn 2, respectively.
The first channel is divided in two parts, each of length $\ell/2$,
called \emph{gates}. Periodic boundary conditions are imposed by letting the second channel close the table
on a circuit.
This constitutes an ergodic billiard \cite{Bunimovich,Bunim05}. We now
add the \emph{bounce--back}
mechanism in the first channel:
when the number of particles in one gate, that are moving toward the other gate,
exceeds a threshold $T$, the horizontal component of the
velocity of those particles is reversed.  The particles coming from the other gate are unaffected by this mechanism and
continue their motion.
Although this dynamics is deterministic, time reversible and phase volumes preserving, it can produce non--equilibrium phase transitions, because the bounce--back 
mechanism implements a sort of negative feedback, that promotes the onset of a non--equilibrium steady state.

For $T \ge N$, the usual ergodic billiard dynamics is realized \cite{Bunimovich,SzaszB}. Thus, for large $N$, the vast majority of time is spent in a 
state in which approximately the same number of particles resides in each urn. That state, like any other, is abandoned to reach still other states, 
with frequency given by the ratio of the respective phase space volumes \cite{Lebowitz1999}, therefore no state is strictly stable. Nevertheless, 
the lifetime of the homogeneous phase rapidly becomes so long, with growing $N$, that such a lack of stability turns physically
irrelevant even at moderately large $N$, consistently with Boltzmann's explanation of his $H$-theorem \cite{Kac59,Cercignani2006}. For $T < N$,
ergodicity also guarantees that,  sooner or later, the threshold will be exceeded in a gate: as long as that event
does not occur, the dynamics is like the ergodic one, which eventually leads to a stationary homogeneous state. When the threshold is exceeded, the standard dynamics is interrupted by
the bounce--back. As evidenced in \cite{CCMRR2020}, for large $N$ and sufficiently
small $T/N$, a larger concentration of particles in one urn leads to an increased frequency of activation of the bounce--back mechanism in the adjacent gate, while particles can flow in from the other urn, incrementing the effect. As a consequence, one urn gets depleted of particles, while the other urn increases its population, until a steady state is reached in which the flow of particles per unit time in the two directions equalize. In this scenario, a microscopic fluctuation suffices to trigger the transition, even when starting from the homogeneous phase.
In this work, the phenomenology is much richer: the second channel allows particles to flow freely, contrasting 
the trend toward inhomogeneous states. Both homogeneous and inhomogeneous states
can thus be realized, and the latter support stationary self sustained currents, as in a battery.

Each half of
the table is now made of one urn, of the adjacent gate and
of the adjacent semi--channel of length $\ell'/2$ of the second channel,
cf.\ Fig.~\ref{fig:model}. Letting $N_1$ and $N_2$ be the number
of particles in the two halves, with $N=N_1 + N_2$, we define the \emph{mass spread} by
\begin{equation}
\chi = \left| N_1 - N_2 \right|/N \label{mass}
\end{equation} 
For simplicity, we define
the net \emph{current} by taking the absolute value of the difference of the number of particles coming from opposite directions and
crossing the vertical line separating the two gates, and by then dividing this quantity by the elapsed time $t$ \cite{Spohn91}.
%and we take the long time limit.
Namely, let $n_{12}(t)$ and $n_{21}(t)$ denote the number of particles that cross, during the time interval [0, t], the vertical line separating the two gates in the direction from urn 1 to urn 2 and in the opposite direction, respectively. The net current flowing in a given channel is thus given by the ratio 
\begin{equation}
\frac{\left| n_{12}(t)-n_{21}(t)\right|}{t} \label{netJ}
\end{equation}
In the large $t$ limit, such a discretely defined current settles to a stationary value related to the asymptotic billiard current. Due to the symmetry of the model, it is irrelevant whether the net current is positive or negative.

The model has been simulated as follows. Our numerical algorithm updates, at each time step, the position of all particles by moving them along straight lines, in the direction of their velocities, over a distance $v\ \delta t$. Here $\delta t$ denotes the time interval between two consecutive collisions of the particles with the physical boundaries of the table and also with the fictitious vertical lines marking either the boundary of each gate with the adjacent urn or the junction between the two gates in the first channel. Elastic reflections at the boundaries of the billiard table are implemented\footnote{The direction of the inward normal at the junction points between the urns and the channels depends on the origin of the colliding particle. Namely, a junction point is considered as belonging to the urn or to the channel if the incoming particle is originally located, respectively, in the urn or in the channel.}.
The initial datum is chosen by fixing arbitrarily the number of particles inside each urn and selecting their positions and velocities at random with uniform distribution. 
Nevertheless, the values attained in the steady state by the observables in Eqs. \eqref{mass} and \eqref{netJ}  resulted insensitive to the initial datum. In particular, in all our simulations we verified that, for any value of the parameters of the model, the same stationary values of the mass spread and the net current are numerically reached by starting both from $\chi(0)=0$ and from $\chi(0)=1$.

\begin{figure}
%\centering
\begin{picture}(100,145)(90,0)
\put(0,0){
\includegraphics[width = 0.55\textwidth, height=0.3\textwidth]{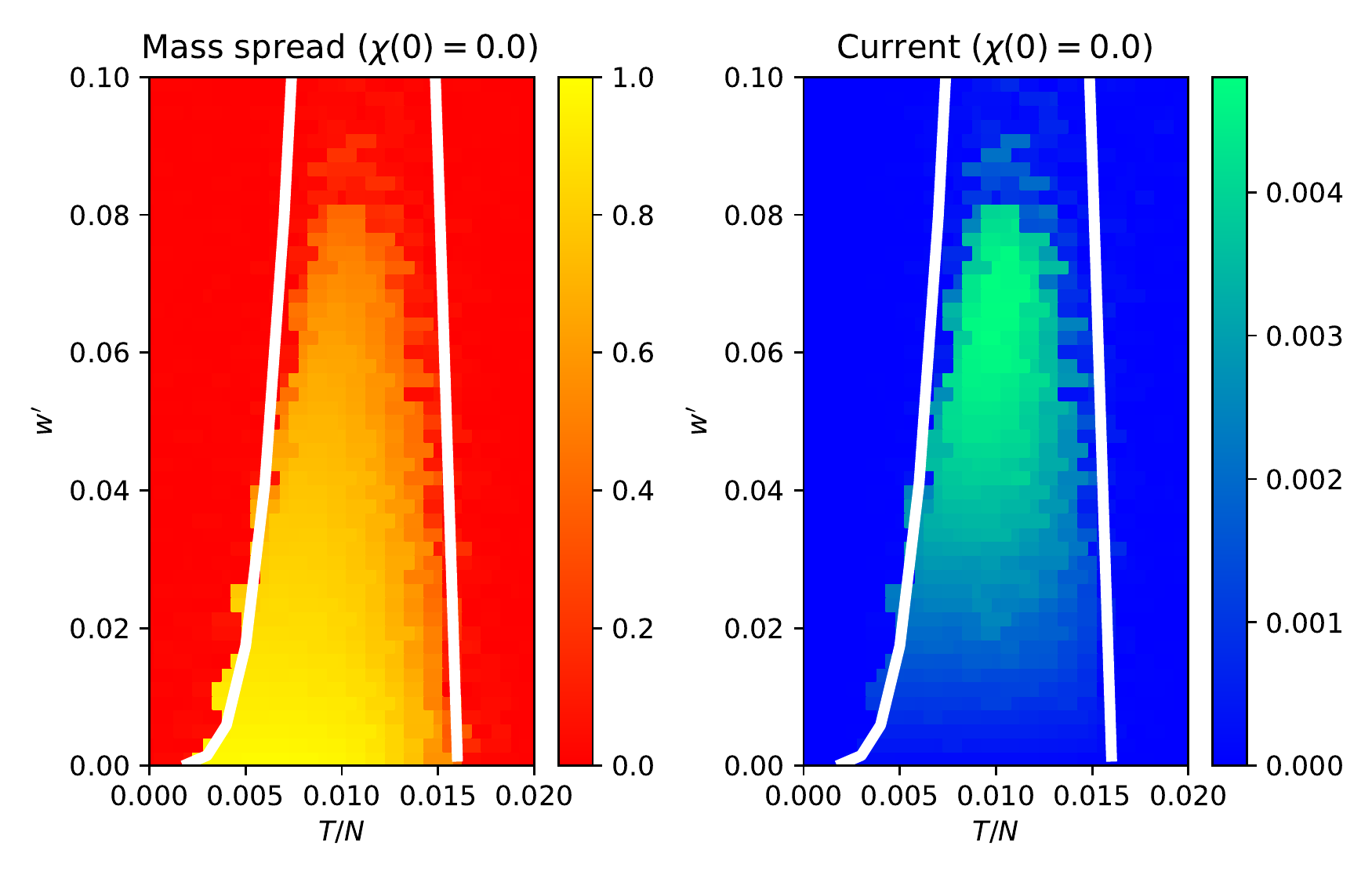}
}
\end{picture}
\caption{Stationary values of the mass spread $\chi$ (left panel) and
net current (right panel) for $N=10^3$, $w=0.3$, $r=\ell=\ell'=1$. The initial condition yields $\chi(0)=0$.
In the left panel, the red pixels denote the homogeneous phase, whereas the other pixels refer to the inhomogeneous phase.
The white lines mark the theoretical boundary between homogeneous and inhomogeneous steady states (see Eq. \eqref{transition}).
}
\label{fig:numris}
\end{figure}

In Fig.~\ref{fig:numris}, we have $N=10^3$, $w=0.3$, $r=1$, $\ell=1$, $\ell'=1$,
and varying values of $T/N$ and $w'$. Initially, $N/2$ particles lie in each urn, while the channels are empty.
Positions and velocities are taken at random with uniform distribution.
The stationary values of $\chi$ and of the net current are computed averaging post--collision data, namely,
right after the
collision of each particle with the walls of the table. Simulations last $10^8$ collisions,
corresponding, on average, to $10^5$ collisions per particle.

%\begin{figure}[h!]
%%\centering
%\includegraphics[width = 0.35\textwidth]{figN-ccrr_circ03.pdf}
%\caption{Behavior of the parameter $\Phi_0$ as a function of $T/N$ for $N=10^3$, $w=0.3$, $R=ell=\ell'=1$, $w'=0.02$.}
%\label{fig:criterion}
%\end{figure}

For $T/N$ in an interval that depends on $w'$, and for small $w$ and $w'$, an inhomogeneous phase is observed together with
a stationary net current flowing in the circuit. In particular, Fig.~\ref{fig:numris} shows that for $w'$ below a certain critical value, 
three different regimes are produced by variations of $T/N$, corresponding to two non--equilibrium phase transitions.
The agreement between the theoretical solid white lines and the numerical results is imperfect because our theoretical calculations rely on probabilistic arguments, which are justified by the ergodic hypothesis, in the large
$N$ and small $w,w'$ limits. Hence, the theory better describes the simulations if $N$ grows.
In Ref.~\cite{CCMRR2020}, where only the first channel is present, the growth of $N$ produces only one interface between homogeneous and inhomogeneous phases, that occurs at a specific $T/N$ value, for fixed geometrical parameters.
%Differently, the model of  allows only one transition and no steady state currents. That paper shows that $T/N$ below
%a given value  leads to frequent bounce--back events, hence to polarized states, and that such states turn stable in the $N \to \infty$
%limit, thanks to the divergence of the recurrence times, cf.\ {\em e.g.}\
%Figs.\ 2 and 3 of Ref.\cite{CCMRR2020}. The point is that, below a certain value of $T/N$, the bounce--back mechanism
%almost stops particles flowing from the most populated urn to the other, while it allows the opposite motion. As $N$ grows,
%this effect becomes sharper and sharper, leading to the phase transition.
The second channel results, instead, in a more complex phenomenology, because the free motion of particles passing through it tends to equilibrate $N_1$ and $N_2$. Therefore, two contrasting mechanisms are at work, and their interplay, between $w'$ and $T/N$ in particular, determines the steady state.

\begin{figure*}
    \centering
    \begin{subfigure}[]
        {\includegraphics[width=0.38\textwidth]{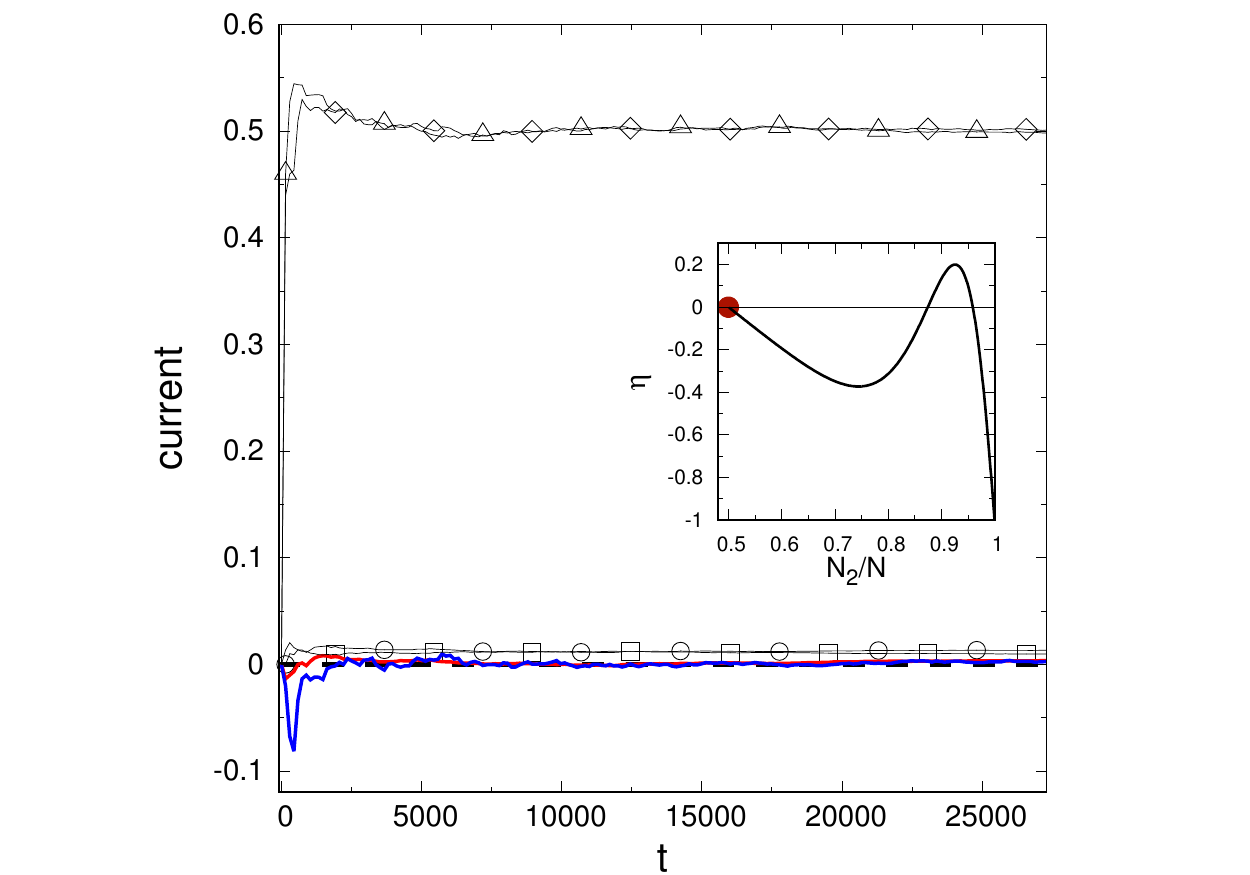}}
    \end{subfigure}
    \hskip -1.5 cm
    \begin{subfigure}[]
        {\includegraphics[width=0.38\textwidth]{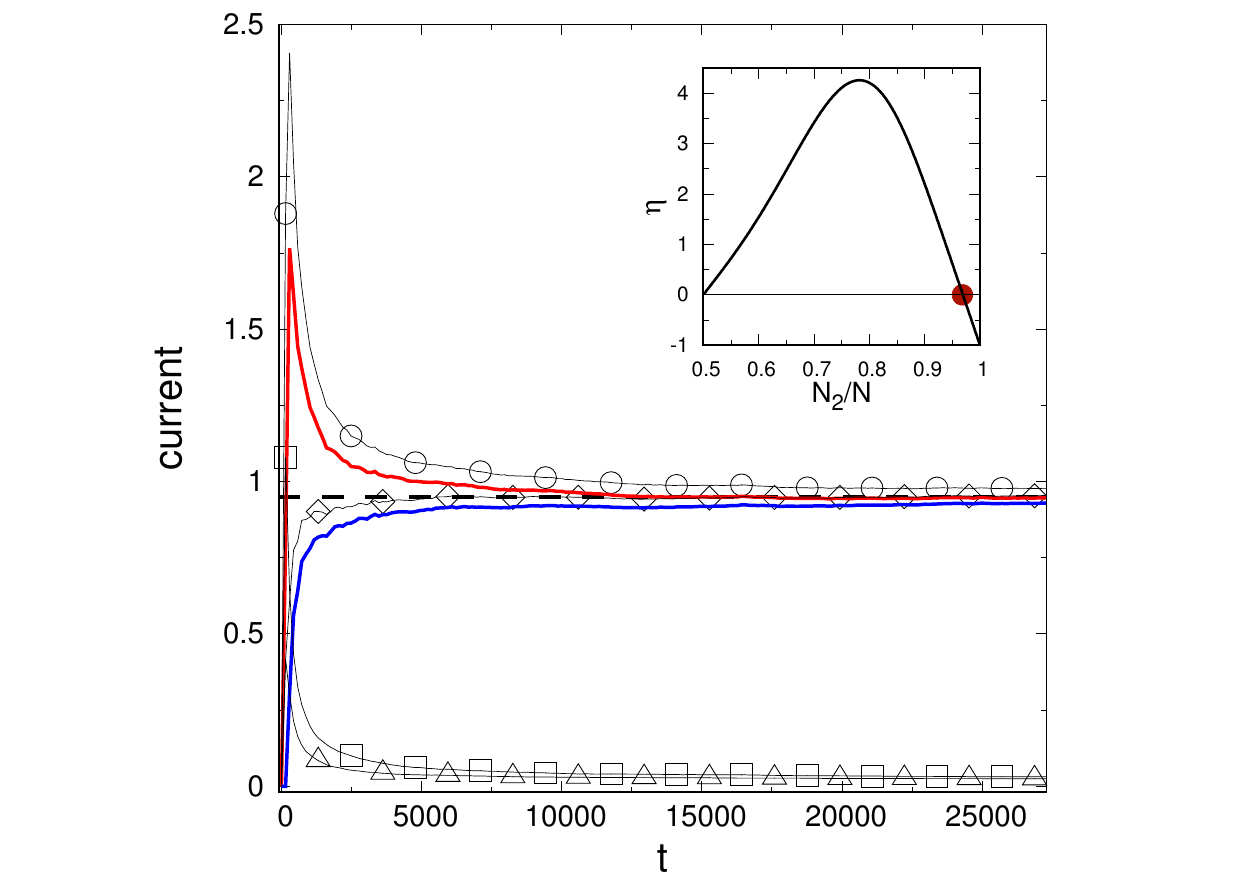}}
    \end{subfigure}
    \hskip -1.5 cm
    \begin{subfigure}[]
        {\includegraphics[width=0.38\textwidth]{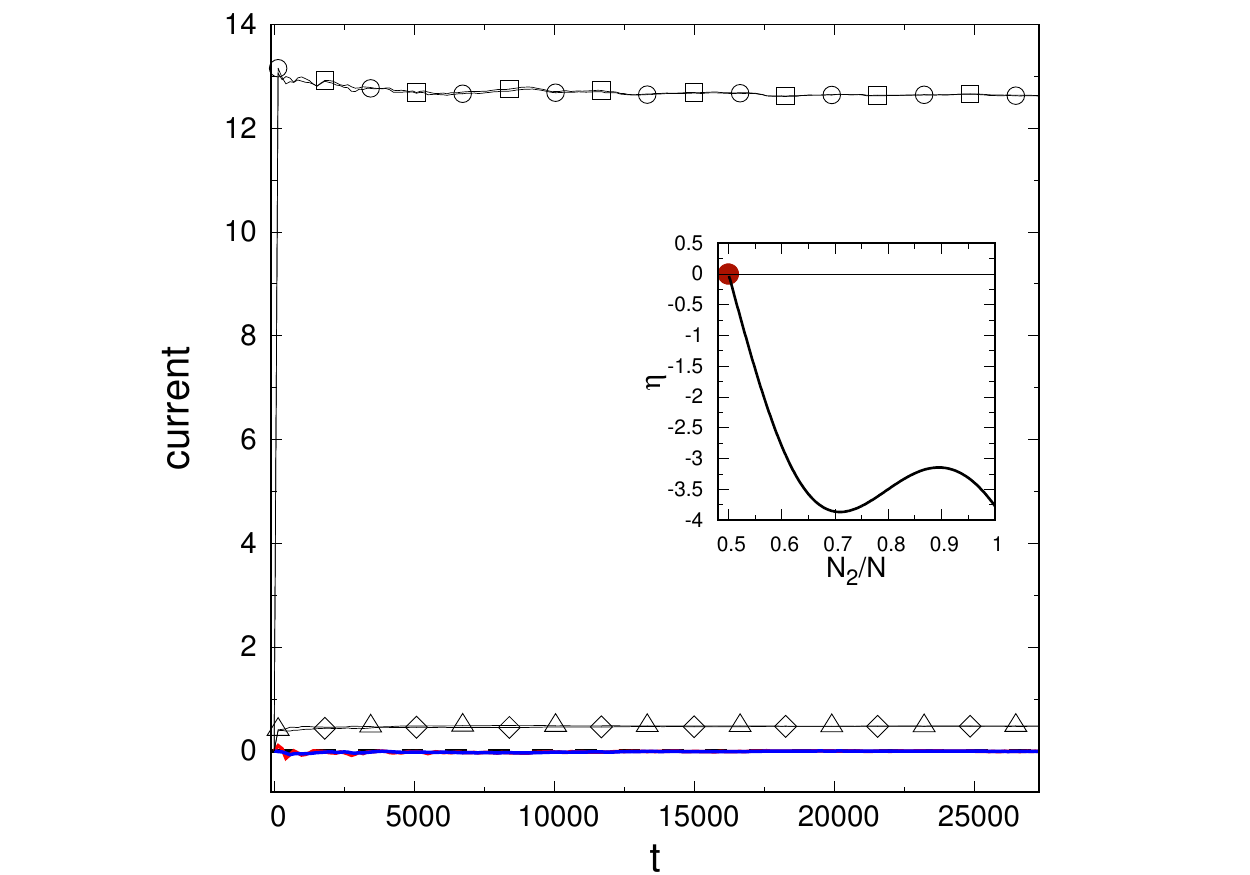}}
    \end{subfigure}    
    \caption{Currents as functions of time, for $r =\ell = \ell' = 1$, $0.01 = w'\ll w = 0.3$, $N = 10^3$, and initial datum such that $\chi(0)=0$. For $T = 2$ (panel (a)) and $T = 18$ (panel (c)), a homogeneous steady state with zero net current is reached. For $T = 7$ (panel (b)), a stationary net current with $N_1\ne N_2$ arises Disks (squares) represent 
numerically computed flows out of urn 1 (urn 2) in channel 1; triangles (diamonds) represent flows out of urn 1 (urn 2) in channel 2. 
Black dashed lines are the theoretical values (obtained from the probabilistic model) of the stationary currents;
solid red (blue) lines denote the net currents in the first (second) channel. The parameter $\eta$ is plotted in the insets as a function of $N_2/N$, while the red disks indicate the stable state reached by the deterministic dynamics.}\label{fig:comp05}
\end{figure*}

\begin{figure}[h]
\includegraphics[width = 0.4\textwidth]{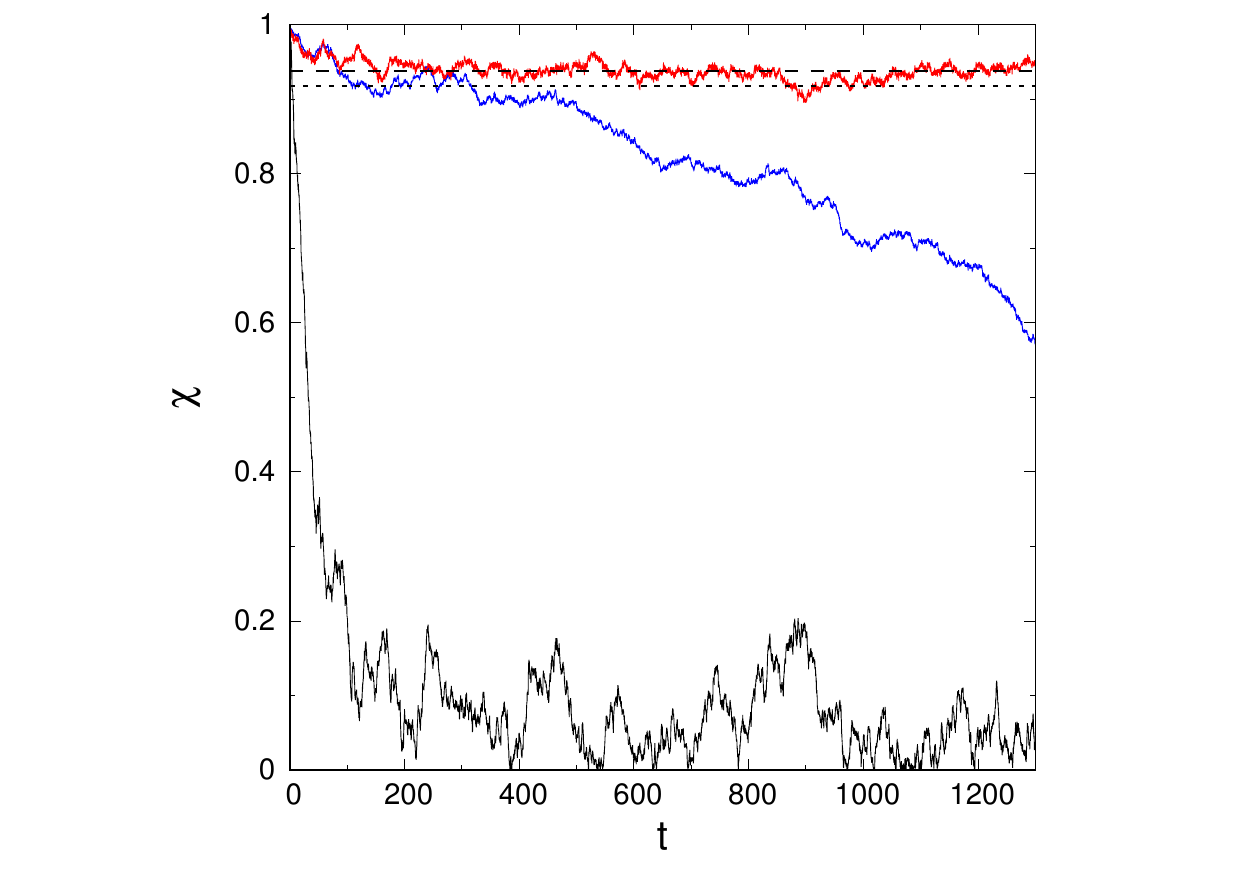}
\caption{Mass spread with the same values of the parameters and of the threshold considered in Fig.~\ref{fig:comp05}: homogeneous and inhomogeneous states are both stable (blue curve, $T=2$); only the inhomogeneous state is stable (red curve, $T=7$); only the homogeneous state is stable (black curve, $T=18$). The dotted and dashed lines indicate the theoretical values of the mass spread for the inhomogeneous states at $T=2$ and $T=7$, respectively. The initial datum is such that $\chi(0)=1$.}
\label{fig:comp06}
\end{figure}

In fact, higher $T/N$ values make the bounce--back mechanism less likely, hence particles
flow more easily through the first channel, while lower values make particles more likely to bounce back.
Flow through the second channel decreases or increases when $w'$ does. 

Figure~\ref{fig:comp05} shows how two different phase transitions can be encountered. 
For $T=2$ (panel (a)), we are in the left region of Fig.~\ref{fig:numris}. Here, a homogeneous phase arises, because 
in the first channel particles frequently bounce back, making left and right flows vanish, but $N_1$ and $N_2$ equalize thanks to the second channel. 
For $T=18$ (panel (c)), in the region to the right of Fig.~\ref{fig:numris}, the first channel allows particles to flow  
almost freely, with a 0 net current, while just a few particles cross the second channel, because much smaller than the first: $w' = w/30$. 
For $T=7$ (panel (b)), we fall in the centre of 
Fig.~\ref{fig:numris}, where an inhomogeneous state, characterized by $N_1\neq N_2$ and by a stationary net current, persists longer than our simulations. The reason is that
the bounce--back phenomenon is only partly mitigated by the flow through the second channel. As the net current in the first channel flows \textit{uphill}, 
{\em i.e.}\ against the population gradient, and \textit{downhill} in the second channel, the first channel acts like an \textit{emf}. 
The numerical results illustrated in Fig.~\ref{fig:comp05} agree with excellent numerical accuracy with the theoretical prediction (black dashed lines) discussed in Sec. \ref{sec:sec2}.
Stationary uphill currents in presence of a non--equilibrium phase transition have been previously observed for stochastic dynamics in 
\cite{CdMP2,CdMP3,CGGV18};
an analogous behavior has also been identified in \cite{CC17} for locally perturbed zero--range processes.
In these cases, uphill currents stem either from a local inhomogeneity in the jump rates, or from the non--equilibrium coupling of the bulk dynamics with external reservoirs, that breaks detailed balance. Our deterministic conservative dynamics accounts, instead, for the work done by 
%a sort of Maxwell's demon: 
the bounce--back mechanism.
This phenomenology can be understood introducing a variation of the probabilistic model of Ref.~\cite{CCMRR2020}, that agrees with our
deterministic dynamics in the large $N$ and small $w,w'$ limits. Details can be found in the Appendix.

\section{Theoretical derivation}
\label{sec:sec2}
 \par\noindent
 
Using the uniformity of the distribution of the particles and of their velocities, one first obtains that the number 
of particles in urn 1, say, entering channel 1 (or channel 2) per unit time is given by $N_1 w v/(\pi A)$ (or by $N_1 w' v/(\pi A)$), where 
\begin{equation}
\label{eq000}
\begin{array}{rcl}
A
& \!\!=& \!\!
\pi r^2
-r^2\arcsin\frac{w}{2r}+\frac{1}{4}w\sqrt{(2r)^2-w^2}
\\
& \!\!& \!\!
\phantom{\pi r^2}
-r^2\arcsin\frac{w'}{2r}+\frac{1}{4}w'\sqrt{(2r)^2-w'^2}
\end{array}
\end{equation}
The number of particles leaving urn 1 and successfully crossing the first channel, per unit time, is then
reduced by the bounce--back mechanism to
\begin{equation}
N_1 {w v \over \pi A} {\Gamma[T,N_1 w \ell/(4 A)] \over (T-1)!}
\end{equation}
$
\label{eq020}
\Gamma[y,x]
=
\int_x^\infty t^{y-1}e^{-s}\,\textup{d}s
\;, \ y>0 \,,
$
being the Euler incomplete $\Gamma$ function. 
Thus, in the probabilistic model, the number of particles leaving urn 1 per unit time, and reaching urn 2, minus those going from urn 2 to urn 1 
is given by:
\begin{equation}
\label{eq007}
\eta
=
\frac{N_1v}{\pi A}
\Big[
w \frac{\Gamma[T,\lambda_1]}{(T-1)!}
+
w'
\Big]
-
\frac{N_2v}{\pi A}
\Big[
w \frac{\Gamma[T,\lambda_2]}{(T-1)!}
+
w'
\Big]
\end{equation}
where $\lambda_i=N_i w \ell/(4 A)$, for $i=1,2$. 
Correspondingly, a steady state implies $\eta = 0$,
%Currents measured in this case are shown in
%the second panel of Fig.~\ref{fig:comp05}.
%the difference between the leaking currents in the primary channel
%and the outlet currents in the secondary one. More precisely,
%the stationarity condition reads
%\begin{equation}
%\label{eq010}
%N_1 \frac{w v}{\pi A} \frac{\Gamma[T,\lambda_1]}{(T-1)!}
%-
%N_2 \frac{w v}{\pi A} \frac{\Gamma[T,\lambda_2]}{(T-1)!}
%=
%N_2\frac{w' v}{\pi A}
%-
%N_1\frac{w' v}{\pi A}
%\;,
%\end{equation}
%where
%$\lambda_i=N_i w \ell/(4 A)$, for $i=1,2$
%(see the center panel of Fig.~\ref{fig:comp05}).
an equation that can be solved for $N_2$. 
From Eq. \eqref{netJ} it is immediately seen that the condition of stationarity amounts to the equality between the net current flowing uphill in the first channel and the net current flowing downhill in the second channel, as in a circuit.
Inspection shows that
$N_2=N/2$ is a solution of $\eta=0$ and also that, for certain parameters values, $\eta$ changes
sign in intervals not containing $N/2$. Given its continuity, in those cases in which $\eta$ has more
than one zero in $[N/2,N]$, one may ask which of the steady states of the probabilistic model 
is stable.
%we consider
%the difference between the total number of particles
%leaving urn 1 and urn 2 per unit of time, namely,
%\begin{displaymath}
%\eta
%=
%\frac{N_1v}{\pi A}
%\Big[
%w \frac{\Gamma[T,\lambda_1]}{(T-1)!}
%+
%w'
%\Big]
%-
%\frac{N_2v}{\pi A}
%\Big[
%w \frac{\Gamma[T,\lambda_2]}{(T-1)!}
%+
%w'
%\Big]
%,
%\end{displaymath}
%with $N_1=N-N_2$,
%where the first and the second
%term are the total current exiting urns 1 and 2, respectively.
%Note that the stationarity condition \eqref{eq010}
%is nothing but the equation $\eta=0$.
Given the smoothness of $\eta$, the linear stability is given by the sign of $\left( {\partial\eta}/{\partial N_2} \right)$:
if positive the steady state is unstable, if negative it is stable. 
The points at which this derivative vanishes delimit the domains of stability of different steady states; hence, as a definition of the theoretical transition line, we shall consider the locus of points such that
\begin{equation}
\frac{\partial\eta}{\partial N_2}\biggr\rvert_{N_2=N/2}=0   \quad .\label{transition}
\end{equation}
In other words, we collect the points where the homogeneous solution of the equation $\eta=0$ becomes unstable.

The stability criterion based on the derivative of $\eta$ is illustrated in Fig.~\ref{fig:comp05}.
The inset in the panel (a) of Fig.~\ref{fig:comp05} shows two stable steady states 
for the probabilistic model, but only the homogeneous one is actually observed in the simulations. This is in accord with the initial 
condition being homogeneous. Possible departures from this state, with $N=10^3$, are expected to be extremely rare. The panel (c)
illustrates a case in which the homogeneous state is stable, is the only steady state for the probabilistic model, and is reached with the deterministic dynamics. 
In the panel (b), the homogeneous state is unstable for the probabilistic model, and is not observed in the simulations, despite initially $N_1=N_2$.
While this shows that the probabilistic model describes quite well the currents in the deterministic dynamics, 
%\textcolor{red}{it remains to be clarified why the simulations of the deterministic dynamics succeed to capture only one of the two stable steady states predicted by the theory. The reason stems from fixing a finite value of $N$ in the simulations: because $N<\infty$, the deterministic dynamics always appears to settle on the homogeneous state, if stable for the probabilistic model. In particular, in presence of two stable states, a homogeneous and an inhomogeneous one (see e.g. the inset of panel (a) in Fig.~\ref{fig:comp05}), the latter is either never reached or is quickly abandoned (depending on the initial datum).}
Fig.~\ref{fig:comp06} shows that some difference 
remains at finite $N$. 
In particular it reports the behavior of the mass spread, for the same values of the parameters and of the threshold considered in Fig.~\ref{fig:comp05}, with an initial datum yielding $\chi(0)=1$. The red line shows the convergence to an inhomogeneous steady state, also evidenced in the panel (b) of Fig. ~\ref{fig:comp05}. The black line shows the estabishment of a stable homogeneous state (see panel (c) of Fig. ~\ref{fig:comp05}), in which the mass spread rapidly drops down around 0. Finally, the blue line illustrates the case, highlighted in the panel (a) of Fig.~\ref{fig:comp05}, in which a
homogeneous and one inhomogeneous states are both stable for the probabilistic model. Starting at $\chi(0)=1$, the deterministic
dynamics converges toward and lingers over the inhomogeneous steady state, but then it moves away, eventually converging to the homogeneous state.  Therefore, for our finite $N$, the lifetime of the inhomogeneous state is short, that of the homogeneous state is very long. That is why the latter looks globally
attracting for the deterministic dynamics.
%In Figs.~\ref{fig:comp01} the solid lines correspond to the stable solutions of
%$\eta=0$;  the points represent numerical results.
%In Fig.~\ref{fig:numris} the white solid lines represent the interface between homogeneous and inhomogeneous steady states.

\begin{figure}
\includegraphics[width = 0.27\textwidth]{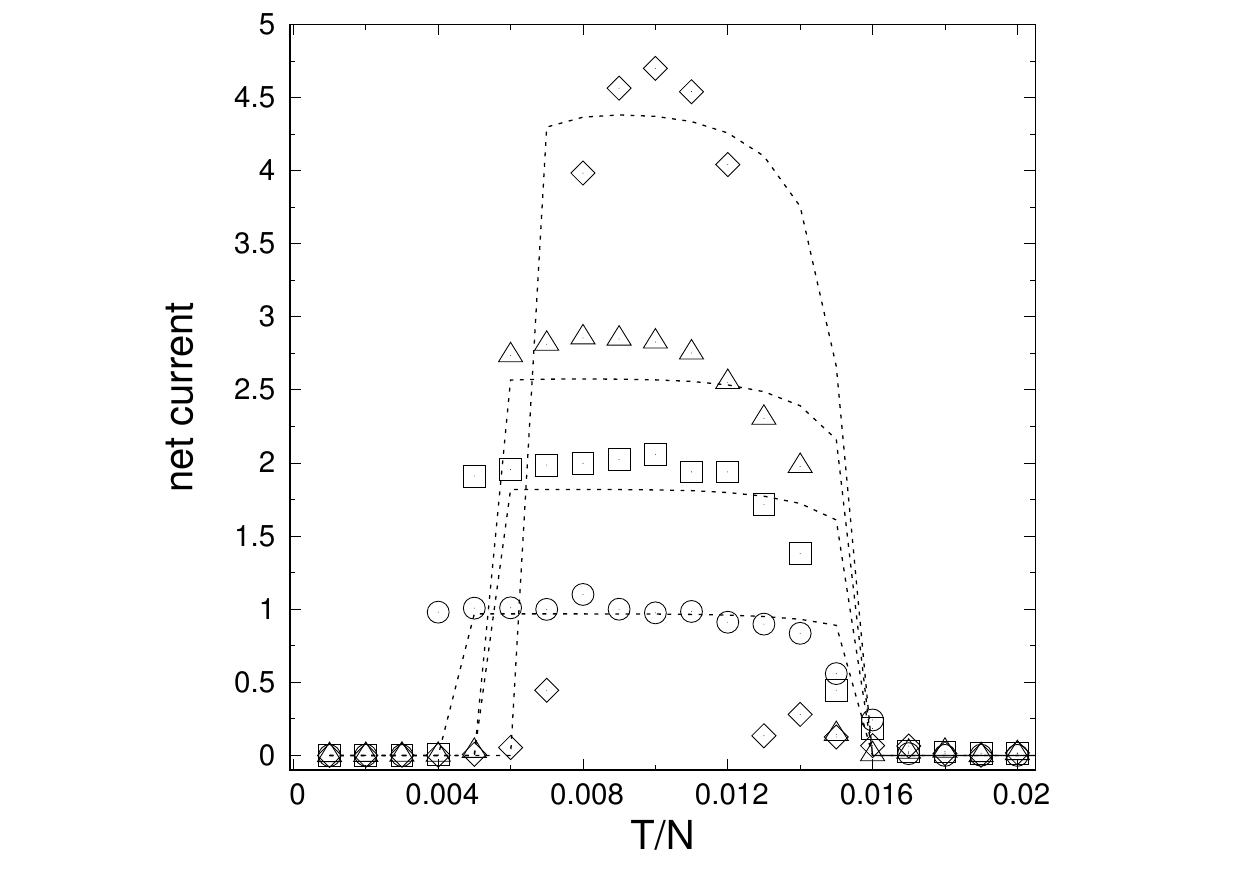}
\hskip -1.2 cm
\includegraphics[width = 0.27\textwidth]{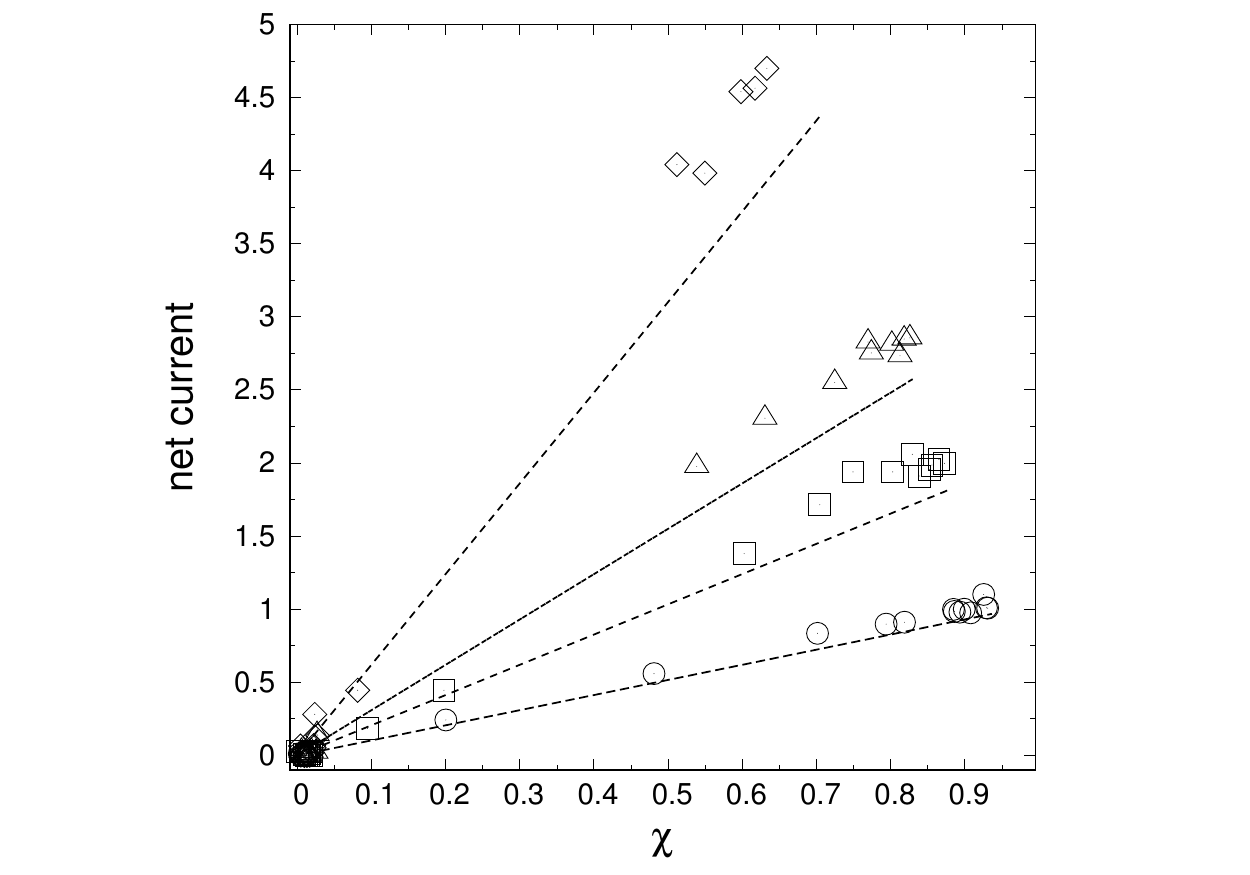}
%\hskip -1.3 cm
%\includegraphics[width = 0.3\textwidth]{fig-ccrr_circ06.pdf}
%\hskip -1.3 cm
%\includegraphics[width = 0.3\textwidth]{fig-ccrr_circ07.pdf}
\caption{Net
currents as a function of $T/N$ (left panel) and of $\chi$ (right panel)
% and $w'$
for
$N=10^3$, $w=0.3$, $r=\ell=\ell'=1$,
$w'=0.0102$ (circles),
$0.0204$ (squares),
$0.0306$ (triangles),
$0.0612$ (diamonds).
%In the two rightmost panels
%$T=6$ (circles),
%$10$ (squares),
%$13$ (triangles).
The dashed lines correspond to the stable solutions of $\eta=0$, see Eq. \eqref{eq007}. }
\label{fig:comp01}
\end{figure}

In the left panel of Fig.~\ref{fig:comp01}, a horizontal slice of Fig.~\ref{fig:numris}, net currents are plotted as
functions of $T/N$ for fixed $w'$.
%(leftmost panels).
%and
%as function of $w'$ for fixed relative threshold $T/N$
%(rightmost panels).
Theoretical predictions and data from simulations are compared, which reveals that for small values of $w'$
the match is good already at $N=10^3$. The right panel of Fig.~\ref{fig:comp01} highlights the {\em battery} phenomenon, with the first channel generating the \textit{emf}. The resulting net current, at fixed $w'$,
is linear with the mass spread $\chi$,  and its slope increases with $w'$, closely following the theoretical prediction.
The linearity is better realized for smaller $w'$, consistently
with the conditions for the applicability of the probabilistic model to the deterministic dynamics.

%\begin{figure}
%\includegraphics[width = 0.27\textwidth]{figN-ccrr_circ06a.pdf}
%\hskip -1.15 cm
%\includegraphics[width = 0.27\textwidth]{figN-ccrr_circ06b.pdf}
%\caption{Left panel:
%current as a function of $\chi$
%for $N=10^3$, $w=0.3$, $r=\ell=\ell'=1$, and
%$w'=0.0102$ (circles),
%$0.0204$ (squares),
%$0.0306$ (triangles),
%$0.0612$ (diamonds).
%Right panel:
%current as a function of the second channel width $w'$,
%for $\chi=0.9$ and the same parameters as in Fig.\ref{fig:comp01}.
%The dashed lines correspond to the stable solutions of $\eta=0$.
%}
%\label{fig:comp02}
%\end{figure}

\section{Conclusions}
\label{sec:sec3}
\par\noindent

We considered a deterministic conservative reversible particle system undergoing a non--equilibrium phase transition, induced by a bounce--back mechanism in one of the channels. Numerical simulations of the deterministic dynamics reveal the existence of a rich phase diagram in the plane $w'$--$T/N$, which includes states with stationary density gradients and stationary currents. Remarkably, the relation between the mass spread and the net current turns out being linear for small values of $w'$, in agreement with the basic tenets of ohmic transport \cite[Chapter 4]{dGM}. The numerical simulations of the deterministic dynamics are also supported by a theoretical analysis based on probabilistic arguments. The match between numerical and analytical results, that strictly requires the $N \rightarrow \infty$ and $w'\rightarrow 0$
limits, is strikingly good even for moderately large N and moderately small values of $w'$. Interestingly, the regime in which the probabilistic model may be meaningfully applied to the deterministic dynamics is relatively easy to achieve in practice. Some relevant open questions still lay ahead; one, in particular, concerns the existence of phase transitions and stationary currents when considering different geometries of the channels and/or of the cavities, or by adding long--range particle interactions. Further challenging mathematical questions concern the investigation of the thermodynamic limit of our model, the relaxation of the particle system toward a nonequilibrium steady state, which could even exhibit anomalous behavior \cite{Ryabov}, and applications to the modelling of physical and chemical kinetics.

\vskip 15pt
\noindent{\bf Acknowledgements.}
%LR is grateful to the University of
%L'Aquila for its generous hospitality, during which part
%of this work has been developed.
%LR acknowledges partial support from partial support from
%MIUR grant Dipartimenti di Eccellenza 2018--2022.
ENMC and MC thank Karlstad University for its kind hospitality.
OR is grateful to the Sapienza Universit\`a di Roma and the
University of L'Aquila for their kind hospitality and
thankfully acknowledges partial financial
support of the GS Magnussons fond.
LR has been partially supported by Ministero
dell'Istruzione, dell'Universit\`a e della Ricerca (MIUR) Grant
No. E11G18000350001 ``Dipartimenti di Eccellenza 2018-
2022''. The authors are grateful to the Laboratorio di Calcolo of SBAI, Sapienza Universit\`a di Roma.

\appendix*
\section{Probabilistic derivation of stationary currents}
\label{sec:app}
\par\noindent
A geometrical argument shows that given $n$ particles uniformly distributed in an urn of area $A$, having fixed speed $v$ and such that the direction of their velocities is uniformly distributed on $[0,2 \pi]$, the typical number of particles that, in a small time interval of length $\delta>0$, leave the urn and enter a gate of width $w$ and length $\ell/2$, is given by 
\begin{equation}
\label{eqapp1}
2 \left(\frac{n}{2 \pi A}\right) w \int_{0}^{v\delta} \arccos{\left(1-\frac{x}{v\delta}\right)} dx=n\frac{w v \delta}{\pi A} 
\end{equation}
We call $p_\delta=w v \delta/(\pi A)$ the probability that one particle in the urn enters the gate in the time interval $\delta$,  and also call $\tau=\ell \pi/(4 v)$ the typical time needed for a particle to cross it. We then introduce a partition of the time interval $\tau$ into segments of length $\delta$. 

The probability that $s$ particles enter the gate in 
the time $\tau$ obeys, in the $\delta\to0$ limit, the Poisson distribution
\begin{equation}
\label{eqapp3}
\frac{\lambda^s}{s!}\,e^{-\lambda}
\;\;\textup{ with }\;\;
\lambda=n\frac{w\ell}{4 A}
\;,
\end{equation}
where $n$ is assumed to be so large not to be sensibly affected by the number of particles entering the gate.
Therefore, the probability that at most $T$ particles enter the gate in
the time interval $\tau$ reads 
\begin{equation}
\label{eqapp4}
P_\tau
=\sum_{s=0}^T\frac{\lambda^s}{s!}\,e^{-\lambda}
=\frac{\Gamma(T+1,\lambda)}{T!}
\;,
\end{equation}
$
\Gamma[y,x]
=
\int_x^\infty t^{y-1}e^{-s}\,\textup{d}s
\;, \ y>0 \, ,
$
being the Euler incomplete $\Gamma$ function.

\begin{figure}
\includegraphics[width = 0.24\textwidth]{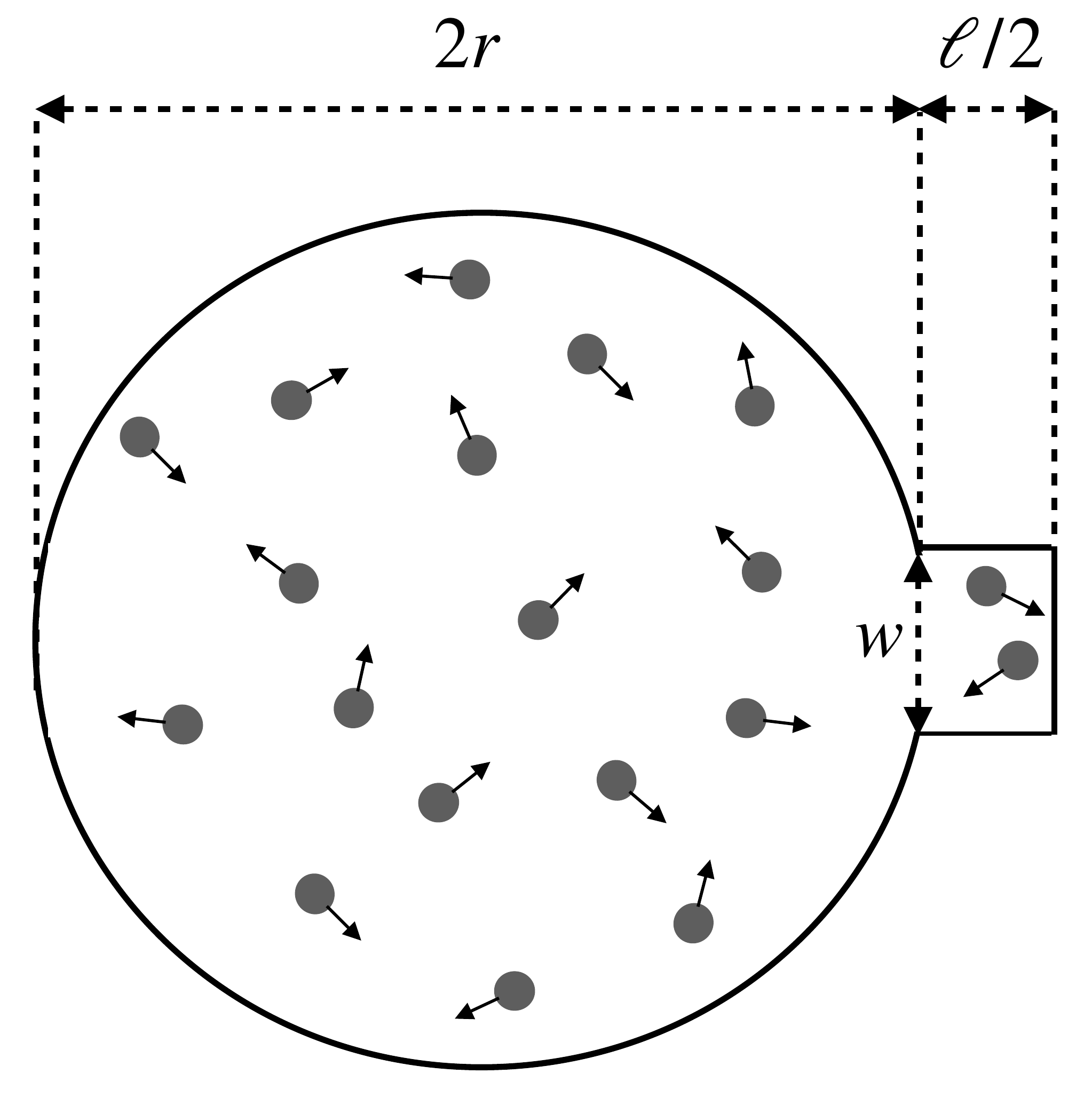}
\caption{The billiard table considered in the numerical simulations.}
\label{fig:app1}
\end{figure}

We now consider a larger time scale of order $t$, and introduce a coarser partition of the latter into segments of length $\tau$. We have that $P_\tau t/\tau$
corresponds to the typical number of events such that, in each time interval $\tau$, there are at most $T$ particles in the gate.
On the other hand, the typical number of particles entering the gate in the time $\tau$, conditioned to the fact that their number be at most $T$, is equal to $\lambda T \Gamma(T,\lambda)/\Gamma(T+1,\lambda)$.
Hence, we find that 
\begin{equation}
\label{eqapp5}
\frac{\lambda}{\tau} \frac{\Gamma(T,\lambda)}{(T-1)!} t= n\frac{w v}{\pi A}  \frac{\Gamma(T,\lambda)}{(T-1)!} t
\end{equation}
yields the typical number of particles that, up to time $t$, successfully exit the gate without being bounced--back by the threshold mechanism \cite{CCMRR2020}.

\begin{figure}
\includegraphics[width = 0.27\textwidth]{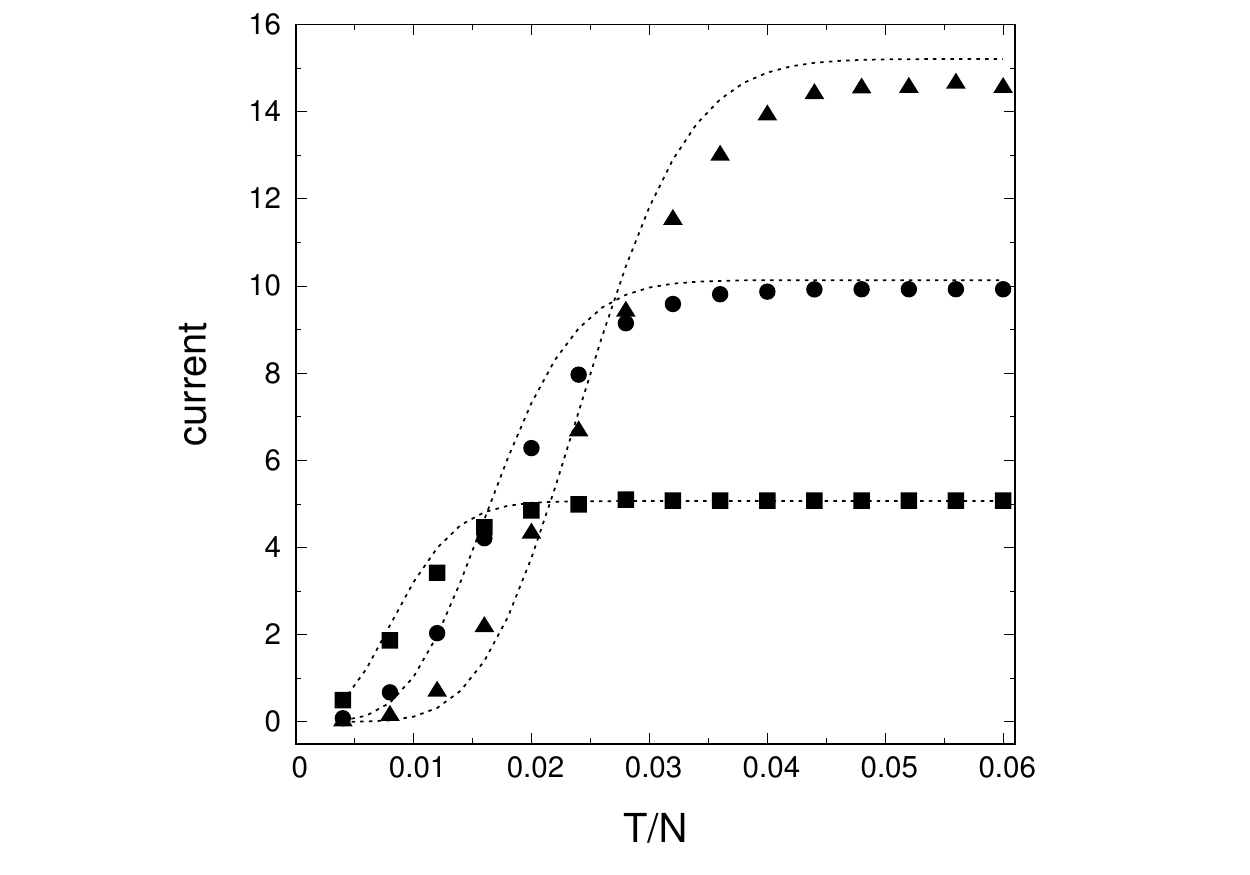}
\hskip -1.15 cm
\includegraphics[width = 0.27\textwidth]{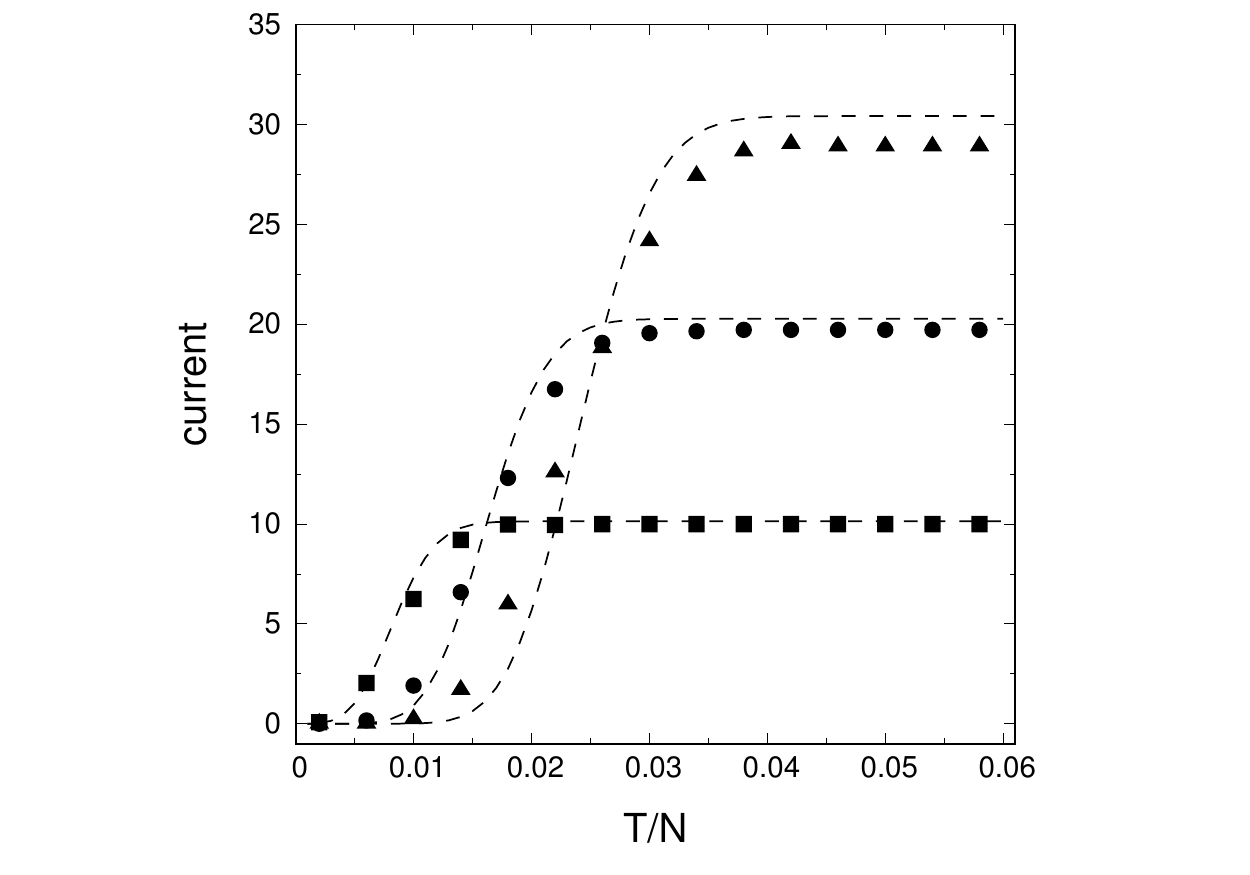}
\caption{Stationary
currents as functions of $T/N$, with $N=500$ and $N=1000$ (left and right panel, respectively),
% and $w'$
for $r=\ell=1$, and
$w=0.1$ (squares),
$0.2$ (circles),
$0.3$ (triangles), numerically obtained from the simulation of the deterministic dynamics.
Dotted and dashed lines correspond to the theoretical values of the current with $N=500$ and $N=1000$ respectively, obtained from Eq. \eqref{eqapp5}. }
\label{fig:app2}
\end{figure}

\begin{figure}
\includegraphics[width = 0.27\textwidth]{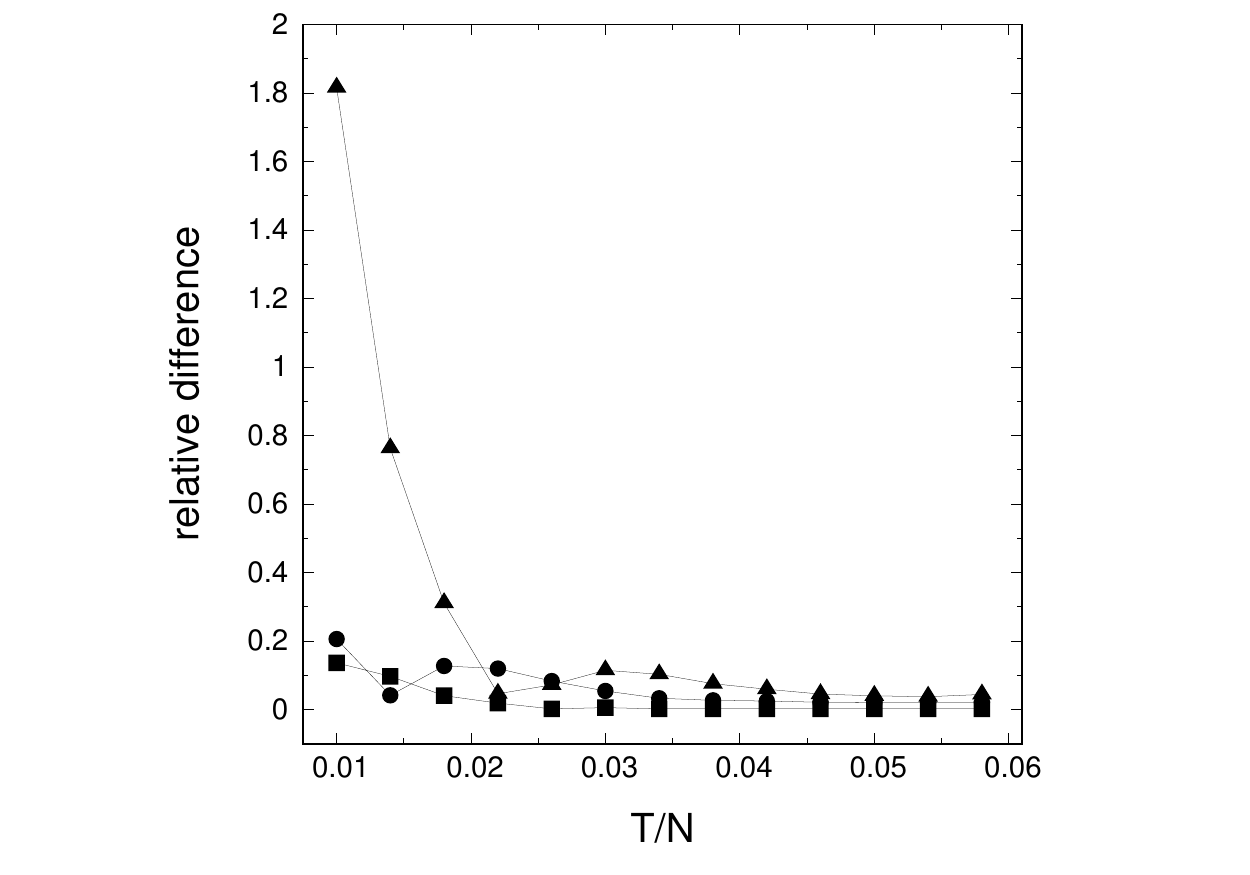}
\hskip -1.15 cm
\includegraphics[width = 0.27\textwidth]{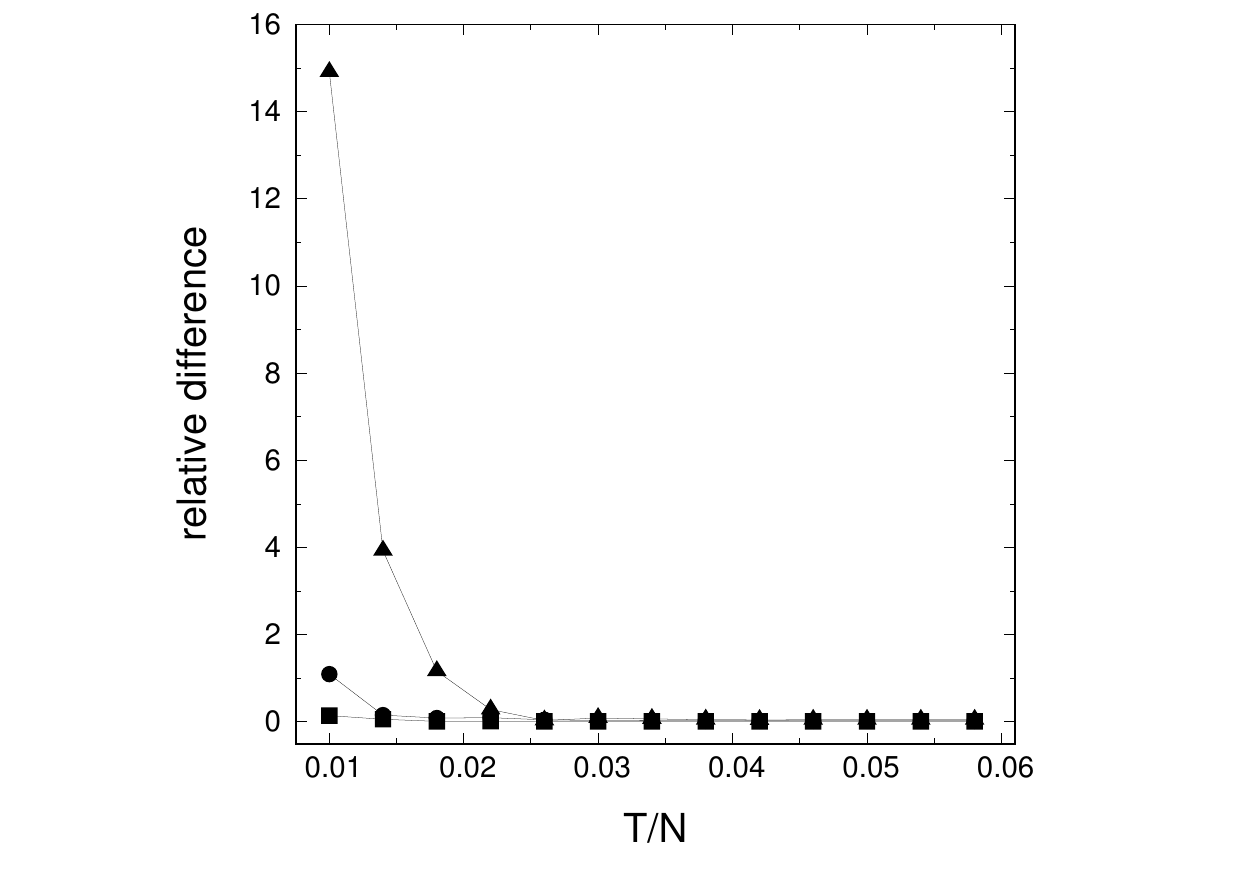}
\caption{Difference between the theoretical and the numerical values of the
currents, divided by the corresponding theoretical values, as a function of $T/N$ with $N=500$ and $N=1000$ (left and right panel, respectively),
% and $w'$
with the same values of the parameters of Fig. \ref{fig:app2}.}
\label{fig:app3}
\end{figure}

\begin{figure}
\includegraphics[width = 0.45\textwidth]{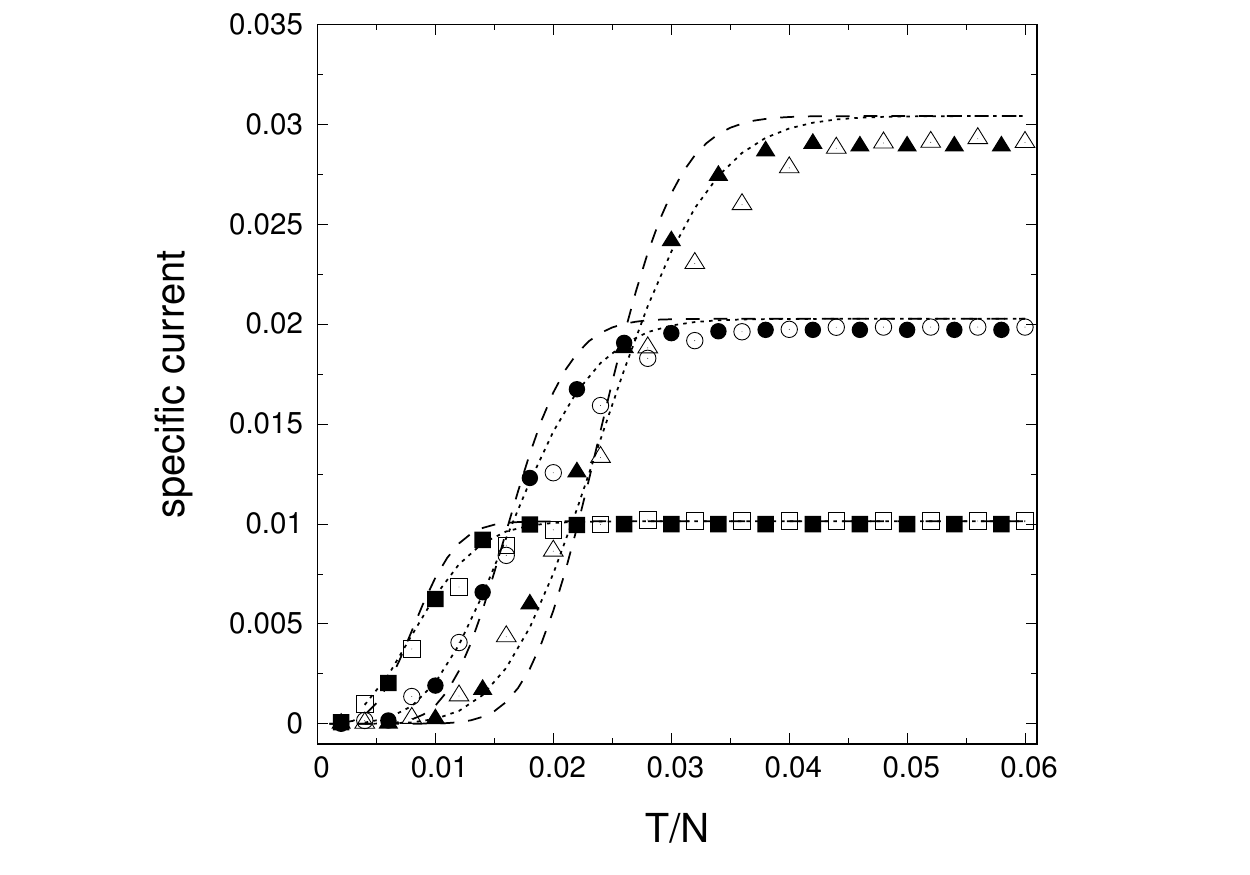}
\caption{Stationary
currents, rescaled by $N$, as functions of $T/N$, with $N=500$ and $N=1000$ (open and and filled symbols, respectively),
% and $w'$
with the same values of the parameters of Fig. \ref{fig:app2}.
Dotted and dashed lines correspond to the theoretical values of the current rescaled by $N$, with $N=500$ and $N=1000$, respectively.}
\label{fig:app4}
\end{figure}

In order to validate the theoretical derivation of the currents, 
based on these probabilistic arguments, we numerically implemented a deterministic 
dynamics similar to the one used in the paper.  
We considered a billiard table made of a single urn and a semi--channel, whose right vertical boundary is an elastically reflecting wall, see Fig. \ref{fig:app1}. Inside the semi--channel, a bounce--back mechanism works as described in the paper.
Positions and velocities of the $n$ particles are initially taken at random, inside the urn, with uniform distribution. 

Such dynamics guarantees that the number of particles in the 
urn is approximately constant, as it equals $n$ minus the (fluctuating)
number of particles in the semi--channel. 
This, hence, permits a direct check of Eq. \eqref{eqapp5}.

The stationary current departing from the urn is defined as the long time limit of the ratio 
of the number of collisions against the right vertical boundary of the semi--channel 
to the elapsed time \cite{Spohn91}. 

Figure \ref{fig:app2} shows, in particular, the comparison between the 
theoretical expression of the current, from Eq. \eqref{eqapp5}, and the numerical 
values of the stationary current obtained from the simulation of the deterministic dynamics. 
The discrepancy between the theoretical and the numerical values of the 
current is highlighted in Fig. \ref{fig:app3}. As visible from Figs. \ref{fig:app2} and  \ref{fig:app3},
the agreement between theory and simulations improves for 
decreasing values of $w$ and for growing values of $T/N$: namely, when the ergodicity 
of the billiard dynamics is restored.

Finally, Fig. \ref{fig:app4} highlights the behavior of the 
\textit{specific current}, i.e., the stationary current divided by $N$, 
for different values of $N$.


\begin{thebibliography}{99}

\bibitem{Kubo91} R. \ Kubo, M. \ Toda, N. \ Hashitsume, {\it Statistical Physics II. Nonequilibrium Statistical Mechanics}, Springer-Verlag, Berlin (1991).

\bibitem{EvMo} G.P.\ Morriss, D.J.\ Evans, {\it Statistical Mechanics of Non--equilibrium Liquids},
Cambridge University Press (2008)

\bibitem{Gall14} G.\ Gallavotti, {\it Nonequilibrium and Irreversibility}, Springer-Verlag, Berlin (2014).

\bibitem{Bertini15} L. \ Bertini, A. \ De Sole, D. \ Gabrielli, G. \ Jona-Lasinio, C. \ Landim, {\it Macroscopic fluctuation theory}, Rev. Mod. Phys. {\bf 87}, 593 (2015).

\bibitem{Lebowitz} G. \ Eyink, J.L. \ Lebowitz, H.\ Spohn, {\it Hydrodynamics of Stationary Non-Equilibrium States for Some Stochastic Lattice Gas Models}, Comm. Math. Phys. {\bf 132}, 253--283 (1990).

\bibitem{Maes} C. \ Maes, K. \ Neto\u{c}n\'{y}, {\it Time-Reversal and Entropy}, J. Stat. Phys. {\bf 110}, Nos. 1/2 (2003).

\bibitem{BePuRoVu}
U.\ Marini Bettolo Marconi, A.\ Puglisi, L.\ Rondoni and A.\ Vulpiani, {\it Fluctuation--Dissipation: Response theory in statistical physics}, Phys.\ Rep.\ {\bf 461} 111 (2008).

\bibitem{Conti2013} L.\ Conti, P.\ De Gregorio, G.\ Karapetyan, C.\ Lazzaro, M.\ Pegoraro, M.\ Bonaldi, L.\ Rondoni,
{\it Effects of breaking vibrational energy equipartition on measurements of temperature in macroscopic oscillators subject to heat flux},
J.\ Stat.\ Mech.\ P12003 (2013). 

\bibitem{CdMP1} M.\ Colangeli, A.\ De Masi, E.\ Presutti, {\it Latent heat and the Fourier law}, Physics Letters A {\bf 380}, 1710--1713 (2016).

\bibitem{BDL10} T.\ Bodineau, B.\ Derrida, J.L.\ Lebowitz, {\it A Diffusive System Driven by a Battery or by a Smoothly Varying Field}, J. Stat. Phys. {\bf 140}, 648--675 (2010).

\bibitem{Kur84} Y.\ Kuramoto, {\it Chemical Oscillations, Waves and Turbulence}, Springer Series in Synergetics. Springer,
Heidelberg (1984).

\bibitem{Wil12} F.\ Wilczek, {\it Quantum time crystals}, Phys. Rev. Lett. {\bf 109}, 160401 (2012).

\bibitem{Zha17} J.\  Zhang, P.W.\ Hess, A.\ Kyprianidis, P.\ Becker, A.\ Lee, J.\ Smith, J.\ Pagano, I.-D.\ Potirniche, A.C.\ Potter, A.\ Vishwanath, N.Y.\ Yao, C.\ Monroe, {\it Observation of a discrete time crystal}, Nature {\bf 543}, 217--220 (2017).

\bibitem{Eich03} R.\ Eichhorn, P.\ Reimann, P.\ H\"{a}nggi, {\it Brownian Motion Exhibiting Absolute Negative Mobility}, Phys. Rev. Lett. {\bf 88}, 190601 (2002).  

\bibitem{Ros05} A.\ Ros, R.\ Eichhorn, J.\ Regtmeier, T.T.\ Duong, P.\ Reimann, D.\ Anselmetti, {\it Absolute negative particle mobility}, Nature {\bf 436}, 928 (2005). 

\bibitem{Muk18} J.\ Cividini, D.\ Mukamel, H.\ Posch, {\it  Driven tracer with absolute negative mobility}, J. Phys. A: Math. Theor. {\bf 51}, 085001 (2018).

\bibitem{CdMP3} M.\ Colangeli, A.\ De Masi, E.\ Presutti, {\it Microscopic models for uphill diffusion}, J. Phys. A: Math. Theor. \textbf{50}, 435002 (2017).

\bibitem{CGGV18} M.\ Colangeli, C.\ Giardin\`{a}, C.\ Giberti, C.\ Vernia, {\it Nonequilibrium two--dimensional Ising model with stationary uphill diffusion}, Phys. Rev. E \textbf{97} 030103(R) (2018).

\bibitem{ACCG19} D.\ Andreucci, E. N. M.\ Cirillo, M.\ Colangeli, D.\ Gabrielli, {\it Fick and Fokker--Planck Diffusion Law in Inhomogeneous Media}, J. Stat. Phys. {\bf 174}, 469--493 (2019).

\bibitem{CCMRR2020}
E. N. M.\ Cirillo, M. \ Colangeli, A. \ Muntean,
O.\ Richardson, L.\ Rondoni,
 {\it Deterministic reversible model of non--equilibrium phase
transitions and stochastic counterpart}
J. Phys. A: Math. Theor \textbf{53},
305001 (2020).

\bibitem{Bunimovich}
L.A.\ Bunimovich, {\it On the ergodic  properties of nowhere dispersing billiards}, Comm.\ Math.\ Phys.\
{\bf 65} 295 (1979).

\bibitem{Bunim05}
L.A.\ Bunimovich, C.P.\ Dettmann,
{\it Open circular billiards and the Riemann hypothesis},
Phys.\ Rev.\ Lett.\
\textbf{94} 100201 (2005).

\bibitem{SzaszB}D.\ Szasz, ed.\ {\it Encyclopaedia of Mathematical Sciences, Volume 101 - Hard Ball Systems
and the Lorentz Gas}, Springer, Berlin (2000).

\bibitem{Lebowitz1999} J.L.\ Lebowitz, {\it Statistical mechanics: A selective review of two central issues}, Rev.\ Mod.\ Phys.\ {\bf 71} S346 (1999).

\bibitem{Kac59} M. \ Kac, {\it Probability and Related Topics in Physical Sciences}, Lectures in Applied Mathematics, Vol. 1, Interscience Publishers (1959).

\bibitem{Cercignani2006} C.\ Cercignani, {\it Ludwig Boltzmann: The Man Who Trusted Atoms}, Oxford University Press, Oxford (2006).

\bibitem{Spohn91} E. \ Spohn, {\it Large Scale Dynamics of Interacting Particles}, Springer-Verlag, Berlin Heidelberg (1991).

\bibitem{CdMP2} M.\ Colangeli, A.\ De Masi, E.\ Presutti, {\it Particle Models with Self Sustained Current}, J. Stat. Phys. \textbf{167}, 1081--1111 (2017).

\bibitem{CC17} E. N. M.\ Cirillo, M.\ Colangeli,  {\it Stationary uphill currents in locally perturbed zero--range processes}, Phys. Rev. E \textbf{96}, 052137 (2017).

\bibitem{dGM} S.R.\ De Groot, P.\ Mazur, {\it Non-Equilibrium Thermodynamics}, Dover Publications, New York (2011).

\bibitem{Ryabov} A.\ Ryabov, {\it Zero-range process with finite compartments: Gentile's statistics and glassiness}, Phys. Rev. E {\bf 89}, 022115 (2014). 

\end{thebibliography}
\end{document}